\documentclass{article}

\usepackage{arxiv}

\usepackage[utf8]{inputenc} 
\usepackage[T1]{fontenc}    
\usepackage{hyperref}       
\usepackage{url}            
\usepackage{booktabs}       
\usepackage{amsfonts}       
\usepackage{nicefrac}       
\usepackage{microtype}      
\usepackage{lipsum}
\usepackage{graphicx}
\graphicspath{ {./images/} }
\usepackage{subfig}
\usepackage{xr} 
\usepackage{wrapfig}
\usepackage{makecell}
\usepackage{adjustbox}

\title{Monitoring War Destruction from Space: A Machine Learning Approach}

\author{
 Hannes Mueller \\
 Institute of Economic Analysis (IAE-CSIC) \\
  08193 Bellaterra, Barcelona, Spain \\
  \texttt{hannes.mueller@iae.csic.es} \\
   \And
 Andre\ Groeger \\
  Department of Economics and Economic History\\
  Universitat Aut\`onoma de Barcelona (UAB) \\
  08193 Bellaterra, Spain \\
  \texttt{andre.groger@uab.es} \\
  \And
   Jonathan Hersh \\
  Argyros School of Business\\
  Chapman University \\
  Orange, CA 92868 USA \\
  \texttt{hersh@chapman.edu} \\
  \And
     Andrea Matranga \\
  Smith Institute for Political Economy and Philosophy\\
  Chapman University \\
  Orange, CA 92868 USA \\
  \texttt{matranga@chapman.edu} \\
  \And
 Joan Serrat \\
  Computer Science Department and Computer Vision Center\\
 Universitat Aut\`onoma de Barcelona (UAB)\\
  08193 Bellaterra, Spain \\
  \texttt{joans@cvc.uab.es} \\
}

\begin{document}
\maketitle
\begin{abstract}
Existing data on building destruction in conflict zones rely on eyewitness reports or manual detection, which makes it generally scarce, incomplete and potentially biased. This lack of reliable data imposes severe limitations for media reporting, humanitarian relief efforts, human rights monitoring, reconstruction initiatives, and academic studies of violent conflict. This article introduces an automated method of measuring destruction in high-resolution satellite images using deep learning techniques combined with data augmentation to expand training samples. We apply this method to the Syrian civil war and reconstruct the evolution of damage in major cities across the country. The approach allows generating destruction data with unprecedented scope, resolution, and frequency - only limited by the available satellite imagery - which can alleviate data limitations decisively.
\end{abstract}



\section{Introduction}
Building destruction during war is a specific form of violence which is particularly harmful to civilians, commonly used to displace populations, and therefore warrants special attention. Yet, data from war-ridden areas are typically scarce, often incomplete and highly contested, when available. The lack of such data from conflict zones severely limits media reporting, humanitarian relief efforts, human rights monitoring, reconstruction initiatives, as well as the study of violent conflict in academic research. One approach has been to use remote sensing to identify destruction in satellite images\cite{Witmer2015}. This approach is gaining momentum as high-resolution imagery is becoming readily available and is updated ever quicker yielding weekly or even daily frequency. At the same time recent methodological advances related to deep learning have provided sophisticated tools to extract data from these images \cite{Lecun2015,Jean2016,Engstrom2017,Yeh2020}. 

 While seminal research has demonstrated the use of automatic classifiers, practical applications have so far been hampered by the high number of false positives generated in real world imagery from urban war zones. These images typically imply heavily unbalanced destruction classes with only a small subset of locations exhibiting severe destruction, which presents big challenges for automatic classification performance. As a consequence, international organizations such as the United Nations, the World Bank, and Amnesty International use remote sensing with \textit{manual} human classification to produce damage assessment case studies \cite{unitar2016,worldbank2017,ai2020}. On the other hand, providers of conflict data for research purposes still rely heavily on news and eyewitness reports which leads to severe data publishing lags and potential biases \cite{Gleditsch2002,Raleigh2010,Sundberg2013,Pettersson2020}. These existing conflict data sets form the basis for our understanding of armed conflict and have been cited thousands of times by academic research. An automated building damage classifier, which has a low rate of false positives on satellite images in unbalanced samples and allows tracking on-the-ground destruction in close to real-time, would therefore be extremely valuable for the international community and academic researchers alike. 

In this article, we demonstrate a new way of combining computer vision techniques, publicly available high-resolution satellite images and a temporal and spatial filtering stage to produce a damage detection methodology which can generate data that are of practical use for practitioners and researchers. We train a type of classifier that has been shown to work well with images -- so-called Convolutional Neural Networks (CNN) -- to spot destruction features from heavy weaponry attacks (i.e. artillery and bombing) in satellite images such as the rubble from collapsed buildings or the presence of bomb craters. We develop and implement a novel data augmentation method, using domain knowledge about the temporal structure of destruction during wartime events, that significantly increases our training data, allowing us to train a building destruction classifier with unprecedented accuracy. We apply this classifier to more than 2 million images from Aleppo city in Syria at multiple dates during the ongoing civil war, and then process the resulting fitted values in a second machine learning stage to build an objective measure of affected areas and an alternative recount of war events. 

The resulting panel dataset provides the first comprehensive and objective measure of building destruction in Syria for the period May 2013 to August 2017 that has been generated by a fully automated method. We demonstrate that this method yields high performance in out-of-sample tests in Aleppo and validate it externally using a separate database of heavy weaponry attacks. Remarkably, the good performance of our procedure generalizes beyond Aleppo which we illustrate in a sample of five other major cities in Syria.

\subsection{Deep learning of destruction}


Training deep learning architectures typically requires large training datasets including thousands of labels which tend to be extremely scarce for our empirical context. The ideal training dataset for destruction classification contains a pixel-wise classification of all damaged and non-damaged buildings across the study area to perform semantic segmentation. This can be achieved by combining building footprint maps with satellite images  \cite{Kahraman16a,Kahraman16b}. When reliable building footprints are not available, as in the case of Syria, an alternative approach is to label damaged buildings through point coordinates. This is done by cropping an image centered around the destruction centroid as a positive sample, and crop negatives randomly from non-overlapping images elsewhere \cite{Gueguen15,Gueguen16}. We follow our own variation of the latter approach leveraging the existence of a large public set of geo-referenced damage labels from the United Nations Operational Satellite Applications Programme (UNOSAT) of the United Nations Institute for Training and Research (UNITAR) \cite{unitar2016}.\footnote{UNITAR/UNOSAT destruction labels were produced from visual inspection of high-resolution satellite imagery and manual annotation of the centroids of destroyed buildings. The data has not been validated in the field.} To get as close as possible to the actual policy application we divide the areas of analysis into patches. We then combine the UNOSAT annotations with these patches to produce labeled data. One problem with this way of generating labels, apart from possible measurement error introduced by human coding, is that buildings have different sizes and, therefore, some UNOSAT labels are surrounded by more visible destruction than others. We address this issue through a second stage in which we smooth the generated data spatially.



\newdimen\figrasterwd
\figrasterwd\textwidth

\begin{figure*}
  \parbox{\figrasterwd}{
    \parbox{.475\figrasterwd}{%
    \subfloat[Full Extent of Aleppo, Syria.]{{
    \includegraphics[width=\hsize]{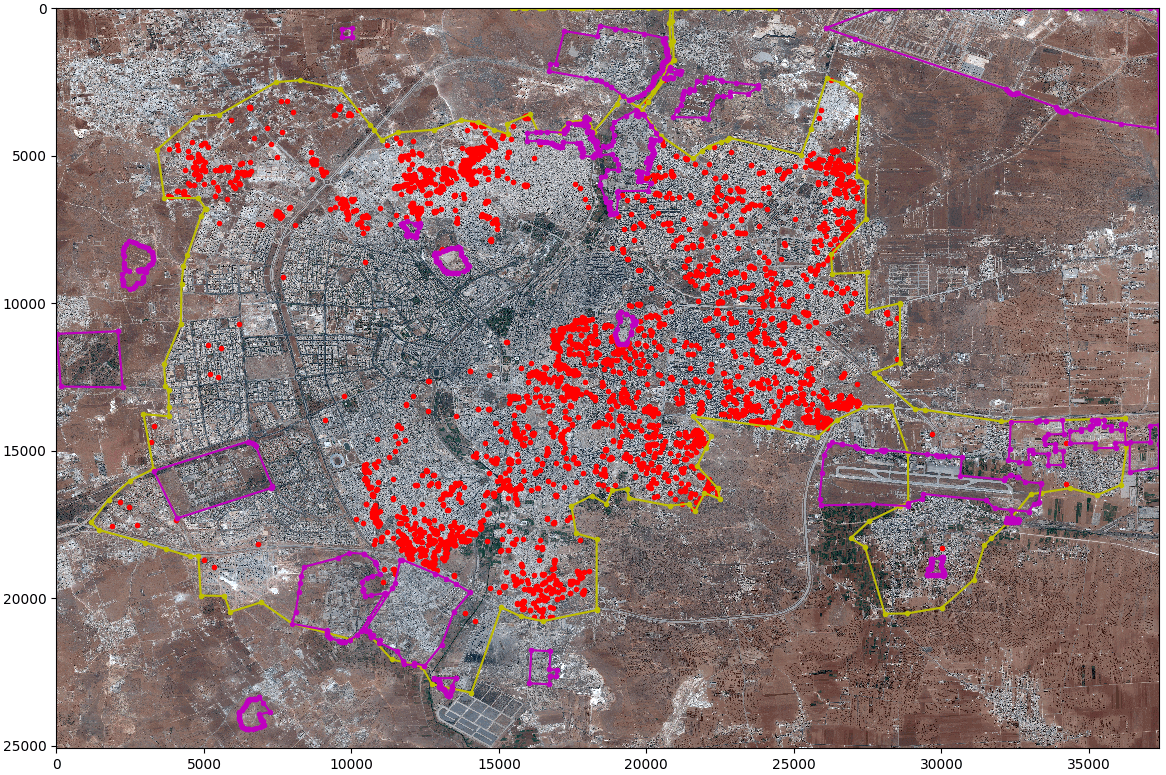}
}}
}
    \parbox{.475\figrasterwd}{%
    \subfloat[Zoom Showing Destruction is Sparse]{{
    \includegraphics[width=\hsize]{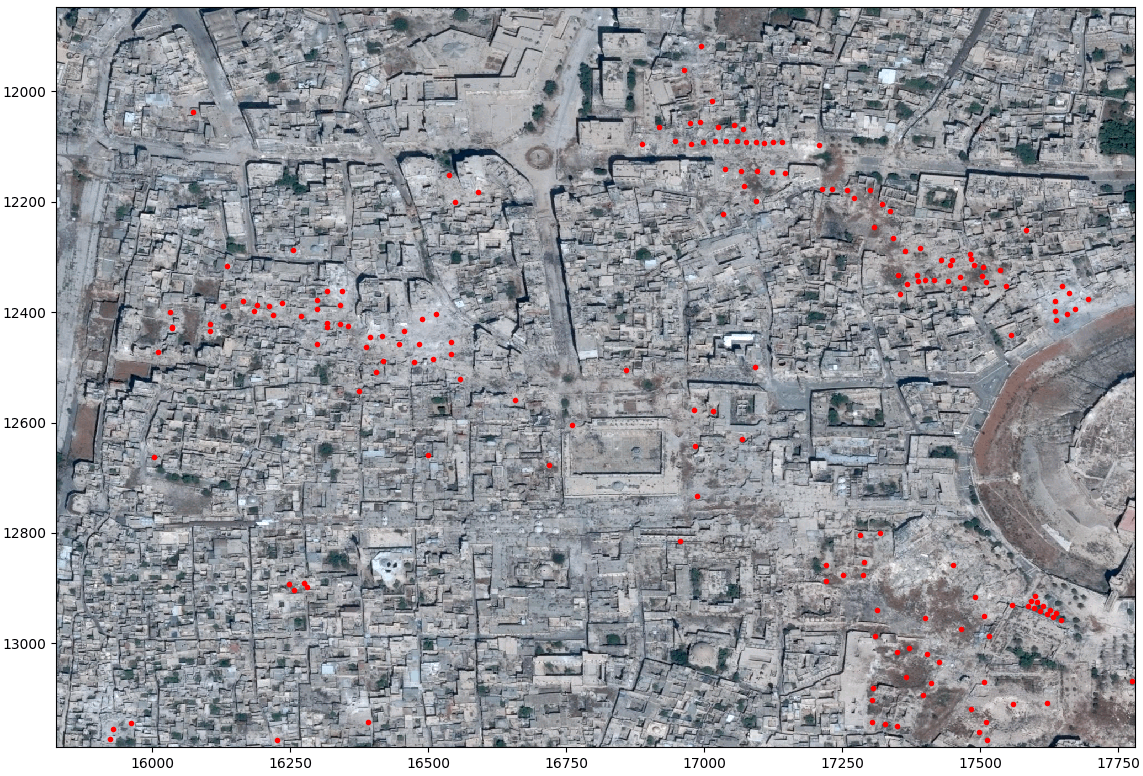}
}}
}}
\caption{Imagery of Aleppo 09/18/2016. Red dots indicate UNOSAT annotations as {\em destroyed}. The regions enclosed in magenta are {\em no analysis zones}, excluded from the UNOSAT damage assessment due to being non-civilian. The yellow lines indicates the boundary of populated areas in Aleppo under analysis. Sources: Google Earth/Maxar Technologies and UNITAR/UNOSAT.}
\label{fig:exhibit_1}

\end{figure*}

A challenge for deep learning of war destruction is related to class imbalances, which stems from the fact that, despite heavy bombardment as in Aleppo, only a small fraction of buildings are destroyed and destruction is highly dispersed across space. This makes deep learning of heavy weaponry destruction more challenging and qualitatively different from destruction triggered by natural disasters, which tends to be spatially clustered \cite{Cooner16,Fujita17,Nex19}. Existing work circumvents this problem by training and testing on samples in which the ratio of positive to negative observations is artificially balanced to 1:1. However, our practical goal is to design a method that is able to evaluate entire cities as in real world applications, i.e. with many more undamaged than damaged locations. In order to do so, we must make meaningful statements about classifier performance on unbalanced and balanced datasets. We view this as particularly important when communicating research to policy makers, who may have inflated expectations based on performance statistics from balanced samples. 

To simulate the real world applications we are interested in, we study the entire populated area of the cities we analyze, including cemeteries, parks and highways (see panel (a) of Figure \ref{fig:exhibit_1}) and we rely on publicly available imagery Google Maps. Most of our sample comes from Aleppo which we use as our main proof-of-concept due to the size of the city and the availability of images and labels. But we also test our methodology in a larger set containing images from five other Syrian cities. Our approach can thus be conducted on any populated area provided that similar high-resolution (i.e. sub-meter) satellite imagery is available.

Our novel data augmentation procedure exploits the fact that reconstruction tends to be largely absent during wartime so that in most cases destruction labels can be extended forward and backward in time. Specifically we assume that any building that is found to be destroyed on a certain date remains destroyed in subsequent images. This process increases the size of our training sample by orders of magnitude to over 2.2 million labeled images of which close to 45,000 show destruction. After training our model using these labels we apply it to scan the whole city of Aleppo repeatedly and produce consistent evaluations of destruction over time. To demonstrate the performance of our method we also conduct rigorous out-of-sample tests and validation exercises. First, we show that we are able to predict destruction in neighborhoods of Aleppo which were completely excluded from the training and testing process as they were left out by UNOSAT's labeling efforts. Second, we validate the timing and location of destruction predictions by means of external heavy weaponry data in an event study setup. Third, by pooling annotations from five additional Syrian cities, we also show that the approach works well on other areas of Syria beyond the city of Aleppo. 

Apart from the practical relevance of providing objective and fine-grained measures of heavy weaponry destruction for humanitarian and development purposes, our approach holds the promise of advancing the academic literature in social sciences which is focusing on understanding the micro-level determinants of violence \cite{Besley2012,Dube2013,Burke2015,Michalopoulos2016,Novta2016,Berman2017}. An additional key application of our method are conflict forecasting systems like ViEWS which rely on spatial dynamics in their forecasts \cite{Hegre2019}. Contrasting our objective measures across cities with war event data will also open completely new avenues for research such as the possible drivers of war reporting, population displacement or the strategic destruction and targeting of neighborhoods.

\subsection{Class imbalances in applications}

Due to class imbalance in destruction prediction tasks, even a small false positive rate (FPR) will result in an unacceptable number of false positive predictions which would yield estimates that are practically useless due to extreme measurement error.\footnote{A simple example illustrates this: suppose you have 100,000 images of which 100 are destroyed. A "low" 15\% false positive rate (FPR) together with a true positive rate (TPR) of 90\% implies that the model will produce 14,985  false positives and 90 true positives, giving a precision of only 0.6\%. In other words, conditional on predicting destruction, such a classifier would be wrong more than 99\% of the time. Note that the same classifier would produce a "high" precision score of 86\% on a 1:1 balanced sample. But this is not representative of the prediction task faced in practical applications.} Therefore, we pay particular attention to designing an approach that accounts for strong imbalances in class distribution in line with the empirical setting of destruction in the Syrian civil war. We show performance metrics against both (unrealistically) balanced and (realistically) unbalanced samples for this prediction task.

Even in a city which suffered as much destruction as Aleppo, only 2.8\% of all images of populated areas contain a building that was classified as destroyed by UNOSAT in September 2016. Figure \ref{fig:exhibit_1} depicts this quite clearly. In the left panel (a) we see the full extent of Aleppo, with all destroyed building annotations depicted as red dots. The right panel (b) zooms into the central area of Aleppo, just East of the historic Citadel, which was heavily attacked. The red dots coincide clearly with patterns of destruction from heavy weaponry attacks in the satellite images, but destruction only affected a minority of buildings even in this heavily affected part of the city.

Several studies have demonstrated the use of computer vision on satellite imagery to identify destruction \cite{Gueguen15,Cooner16,Gueguen16,Kahraman16a,Kahraman16b,Yuan2016,Attari17,Fujita17,Nex19}. Most of this work focuses on highly clustered destruction from natural disasters, which is a qualitatively different problem from our application due to higher class balance. While selected performance results from the literature are encouraging as a proof of concept, they focus on training and then validating on datasets composed of equal numbers of damaged and undamaged images. We argue that performance measures from such artificially balanced samples need to be interpreted with caution when evaluating the benefits of a particular method for real-life applications. Precision performance in scans of entire cities, as in our application, has not been addressed in the literature so far. We emphasize that any useful classifier in our context must be able to detect building destruction in an empirical context where the vast majority of images do not feature destruction.

\begin{figure*}[hb]
\includegraphics[width=\linewidth]{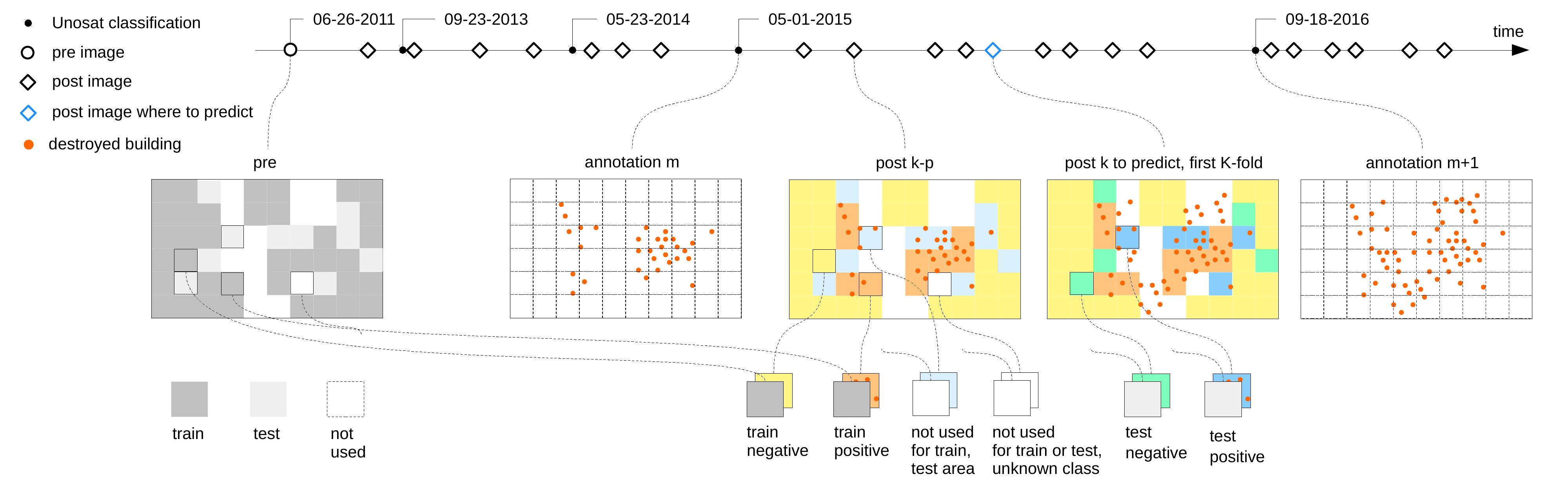}
\caption{Image Sampling and Prediction Process. Timeline shows 23 Aleppo images. The first image, from 06/26/2011 is used as pre image when training the classifier. All other images are used as post images. Images are split into over 95,000 patches which serve as a unit of analysis and are separated into test and training sample before the analysis. Labels for the patches come from UNOSAT annotation dates shown as black dots on the timeline. Annotations are extended forward and backward in time beyond these dates under the assumption that buildings which are destroyed remain destroyed throughout the period of observation and those which are not destroyed were not destroyed before. Patches which are not destroyed at an annotation date but destroyed at a later annotation date have an unknown class. All patches which are not classified as destroyed in the last annotation date are of unknown class (set to missing) after that date.}
\label{fig:exhibit_2}
\end{figure*}

\section{Methods}

\subsection{Data sources}

To train our deep learning algorithm we focus on Aleppo, Syria. The city experienced extensive heavy weaponry attacks during the Syrian civil war. Intense war action started in the second half of 2013 when rebel forces captured large parts of the city and ceased in early 2017 when control was taken back by the Syrian regime with the help of the Russian air force. We use 23 high-resolution satellite images from Aleppo taken between 2011 and 2017 \cite{google_maps}. 


We use the first image from 26 June 2011 as the ``pre'' image as up to that moment the city's building infrastructure had not been affected by the war yet. We define the subsequent 22 pictures as the ``post'' images and combine them with the UNOSAT point annotations which were produced by manually spotting destruction in satellite pictures for four different dates, one each year between 2013 and 2016. These damage annotations for Aleppo (and other Syrian cities) were categorized in three degrees of damage: moderate and severe damage as well as completely destroyed. In our analysis we rely on the destroyed class and exclude the moderate and severe damage labels due to the fact that destruction patterns for these labels were not clearly and consistently visible in the satellite images. To label and process images we divide each satellite picture of the populated area of Aleppo up into over 95,000 patches of images of 64x64 pixels each. These patches are our unit of analysis for training, testing and what we call \textit{scanning} of \textit{dense} predictions where we apply the trained neural network to generate predictions for all patches and times. We do change detection -- i.e. for each patch the pre image is compared to the respective post image of the same patch -- at all 22 later dates. The result is a panel dataset of destruction predictions at the patch level for 22 time periods with approximately 2.1m observations.



Note that we focus on the urban areas of Aleppo as depicted by the area enclosed by the yellow lines in Figure \ref{fig:exhibit_1}. This choice simply stems from the fact that our objective is to detect building damage from heavy weaponry. Including large non-urban areas in this exercise could confound the algorithm to learn distinguishing urbanized versus rural areas only. Purple areas correspond to so-called "no analysis" areas, which have been left out by UNOSAT in their damage classification due to these zones hosting non-civilian buildings. Consequently, these areas are left out in the training process too. We make use of these areas for out-of-sample testing. Sample imagery patches for destroyed areas pre- and post- destruction are presented in Figure S1 and non-destroyed area image patches are presented in Figure S2 in the Supplementary Information.

\subsection{Domain bias and data augmentation}

The computer vision task is to train a convolutional neural network (CNN) to be able to recognize destruction from the visual bands of high-resolution daylight satellite images. This is complicated by both the immense variety of building structures across space, and the changing temporal patterns across images. These problems are known as spatial and temporal domain bias, respectively \cite{sun2016deep}. We deal with this by carefully exploiting the panel dimension of our data: for every patch we have a series of images and labels which help adjusting our training procedure to minimize these biases. 

Our training and testing approach is illustrated in Figure \ref{fig:exhibit_2}. Given the structure of the data, extra care must be taken when splitting the sample for training and testing to avoid overfitting. Standard cross-validation  procedures are not applicable in our context since they could mistakenly mix the training and testing samples along the spatial and longitudinal dimension, i.e. show the network patches from different times, but the same location in training and testing. We therefore use the patch identifier to do a sample split, whereby 70\% of patches are reserved for training and 30\% for testing. Our method classifies the satellite images as destroyed if at least one destroyed UNOSAT annotation is overlapping with the patch at the respective date. 

The size of our training dataset is constrained by the fact that we have 22 post images, but only 4 UNOSAT annotations. To gain more labels we exploit the fact that reconstruction was largely absent in Aleppo during the study period between 2013 and 2017. This allows us to exploit the time dimension of our dataset by using the UNOSAT annotations before and after an intermediate image date to label the patches that do not correspond to an annotation date directly. Consequently, we employ data augmentation by assuming that positive samples at time $t_i$ also remained positives at subsequent times $t_j>t_i$. And conversely, that negative samples at time $t_j$ also had to be negatives at times $t_i<t_j$.

Three cases are then possible for an image which lies between two labels in time as illustrated in Figure \ref{fig:exhibit_2}: 1) the earlier UNOSAT annotation shows destruction and the later patches are labeled as destroyed in line with the no reconstruction assumption (depicted by orange or blue patches), 2) both annotations -- before and after -- show no destruction and we can, therefore, label the intermediate patches as not destroyed as well (yellow or green patches), 3) the early annotation shows no destruction but the later annotation does. In this latter case we do not know exactly when destruction happened and, consequently, we discard these patches at those dates (white patches). All patches that are labeled as not destroyed during the last annotation date are of unknown class after that date.

We solve two problems using this approach. First, we expand the size of our training data set by boosting the number of labels. Second, by including additional time periods in our training sample, we can improve the performance of our classifier to deal with temporal domain bias. Given our strategy of extending labels forward and backward in time it becomes particularly important to verify the ability of our method to approximate the timing of destruction which we verify by means of an event study relying on a completely separate dataset of geolocated bombing events.

\subsection{CNN architecture and two-stage classification procedure}

We use a flexible CNN architecture, which is summarized graphically in Figure S3. Given this structure, we optimize over a set of hyper-parameters to determine the optimal network architecture. Each network uses a series of convolutional blocks. We use convolution because objects in an image correspond to a particular spatial arrangement of pixels and modeling images therefore requires a method that captures the local interactions between input pixels. The idea of convolution is to set neighboring pixels of a location in a relationship to each other through filters which stride over the image and which provide local summaries of the image. 

Each block contains: (i) a convolutional layer, followed by (ii) a rectified linear unit (ReLu) activation function, (iii) a max-pooling layer and (iv) a final dropout layer. The output of the last convolutional block is flattened to a vector and input to two successive fully connected layers followed by a non-linear activation, with a batch normalization layer between them. We explore the space of parameters formed by the number of convolutional blocks, the size of the convolutional kernels, the \textit{stride} of max-pooling, the probability of dropout, the number of units in the fully connected layers and the type of non-linear activation (which can be \textit{ReLU} or \textit{sigmoid}). The optimal hyper-parameters for our architecture are provided in the Supplementary Materials and in the GitHub repository.

The final step of the first stage is to apply the trained model to produce predictions for every patch in every post image as in practical applications of monitoring destruction. In this step we evaluate over 3 million patch images and produce a continuous prediction score for all patches and times in all cities we analyze. The result is what we call \textit{dense} predictions and this forms the raw material for the second stage of our methodology. However, when reporting on performance we will always rely on training on only 70\% of the patches and report performance in the 30\% test sample. 


In the second stage, to exploit the temporal and spatial clustering of destruction, we train a random forest model including two spatial and time lags for each patch to improve precision. The logic behind this is that destruction is not only serially correlated, but also spatially clustered. Indeed, it can be shown in the dense prediction data that both the spatial environment and time lags are extremely informative for whether a patch has been destroyed. We separate this step from the deep learning stage for maximum flexibility and modularity. We use a random forest model to exploit both information on the mean prediction scores around an image and their standard deviation. In the results section we show how this step improves precision. To generate a binary prediction from the random forest model we pick a cutoff that reaches 50 percent recall in the training sample.

\section{Results}

\subsection{Performance}


\newdimen\figrasterwd
\figrasterwd\textwidth

\begin{figure*}[h]
  \centering
  \parbox{\figrasterwd}{
    \parbox{.3\figrasterwd}{%
      \subfloat[Low Precision in Unbalanced Sample.]{{
      \includegraphics[width=\hsize]{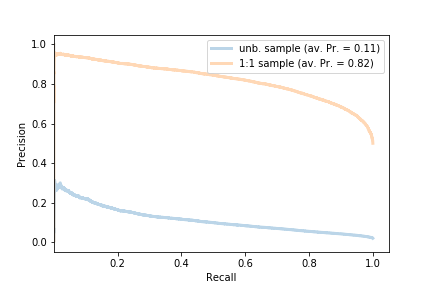}}
      }
    \vskip1em
      \subfloat[Second Stage Precision Improvement.]{{
      \includegraphics[width=\hsize]{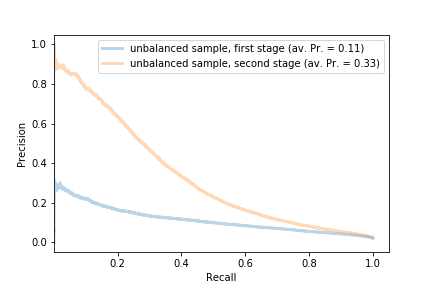}}  
        }
    }
    \hskip1em
    \parbox{.6\figrasterwd}{%
      \subfloat[Patch-Wise Second Stage Destruction Prediction Scores for Aleppo City, Syria.]{{
      \includegraphics[width=\hsize, height=8cm]{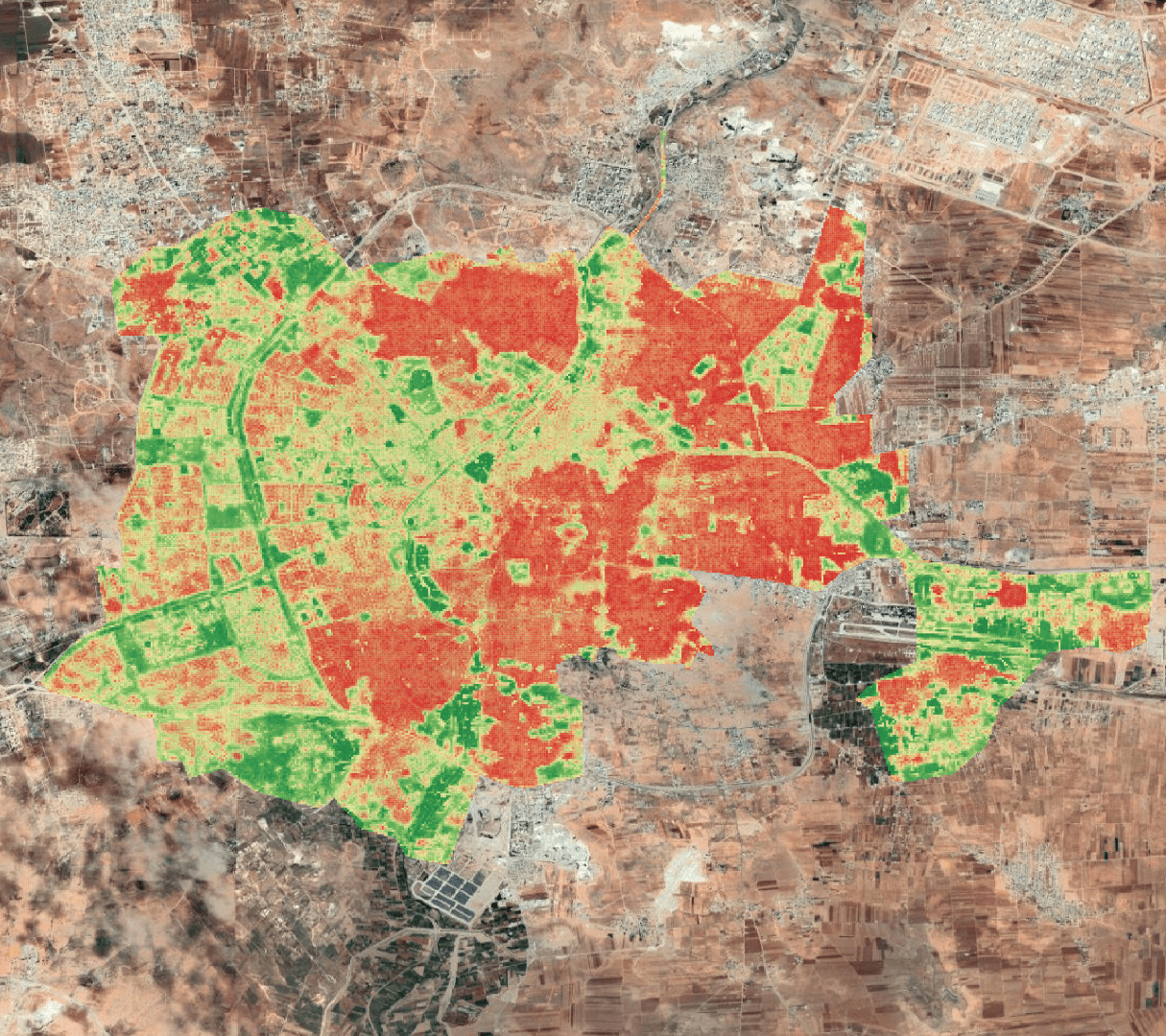}}
      }
    }
  } 
\caption{\emph{(Top Left)} Precision-Recall Curve, Unbalanced Versus Balanced Sample. Reported performance is in the 30\% training sample either by up-sampling the positives to reach a 1:1 sample (orange curve) or by evaluating at the original sample proportions (blue curve). \emph{(Bottom Left)} Precision-Recall Curve, Unbalanced Sample. First Stage Performance Versus Second Stage Performance. Blue curve shows performance after the first stage. Orange curve shows performance after the second stage which uses training of a random forest on temporal and spatial leads and lags in the training sample. \emph{(Right)} Resulting Second Stage Dense Patch-Wise Destruction Prediction Scores for Aleppo City, Syria. Green color indicates low prediction scores, red color indicates high prediction scores. Color bins reflect deciles of fitted values. Sources: Google Earth and author calculations.}
\label{fig:exhibit_3}
\end{figure*}

Our CNN classifier achieves an Area Under the Curve (AUC) of 0.84 in the test sample of the first stage, i.e. with the raw output from the network (see Figure S4). The AUC is a classification performance measure which is not affected by class imbalance in the sample and, therefore, helps us to compare our performance to existing applications in the literature. However, a direct comparison remains partial as several existing papers on damage classification do not report AUC measures. For those that do, the performance of our model is found to be on par or superior to existing studies despite the fact that we use only publicly available images. In what follows, we focus on the precision statistics to illustrate the problem of unbalanced classes in destruction detection applications.


\begin{figure*}[ht]
  \centering
  \parbox{\figrasterwd}{
    \parbox{.33\figrasterwd}{%
      \subfloat[Raw Satellite Image 06/12/2016]{{
      \includegraphics[width=\hsize]{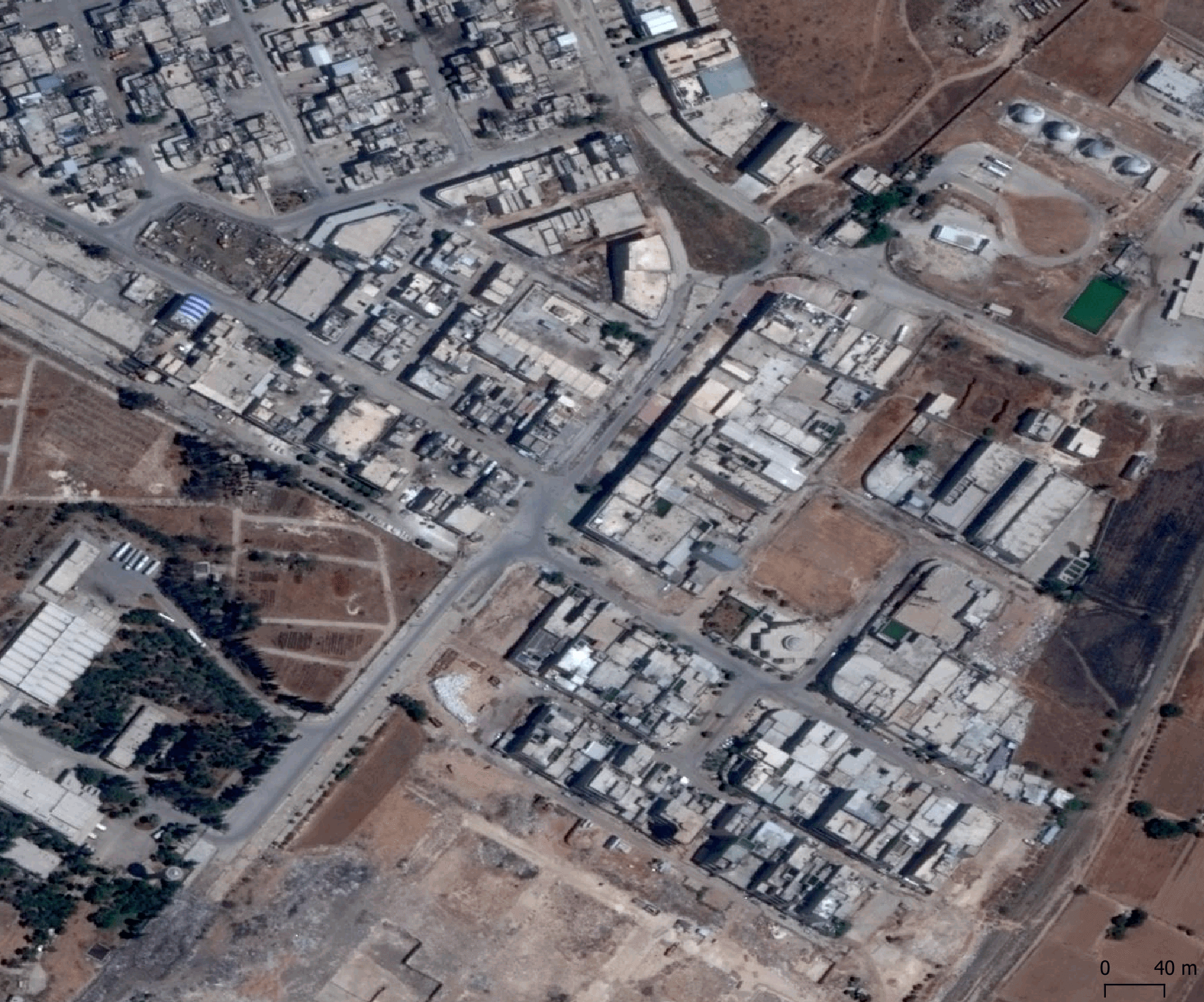}}
    }}
    \parbox{.33\figrasterwd}{%
      {\subfloat[Continuous Prediction Scores 06/12/2016]{{
      \includegraphics[width=\hsize]{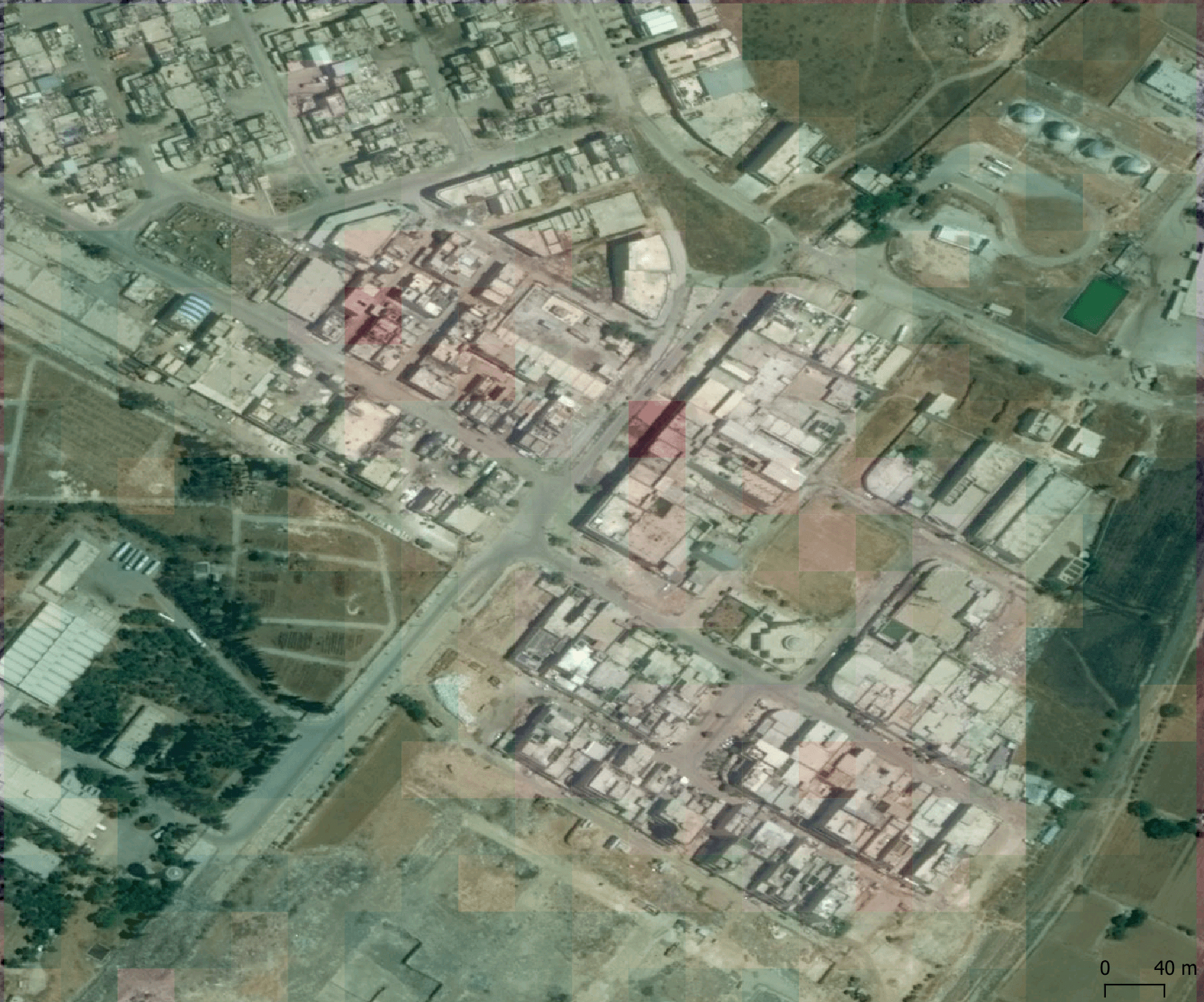}}
    }}}  
    \parbox{.33\figrasterwd}{%
      \subfloat[Binary Predictions 06/12/2016]{{
      \includegraphics[width=\hsize]{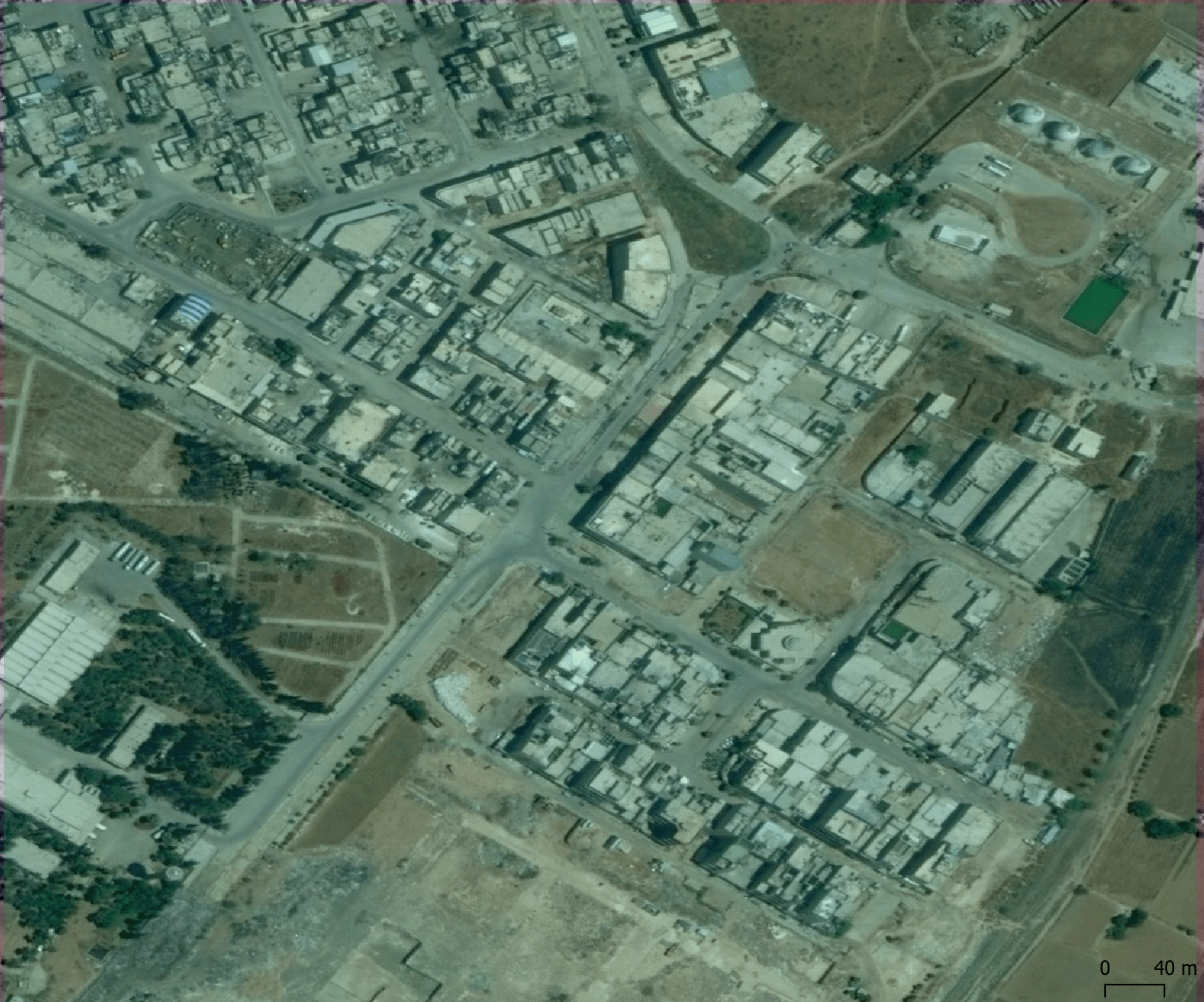}}
    }}
    \parbox{.33\figrasterwd}{%
      \subfloat[Raw Satellite Image 09/18/2016]{{
      \includegraphics[width=\hsize]{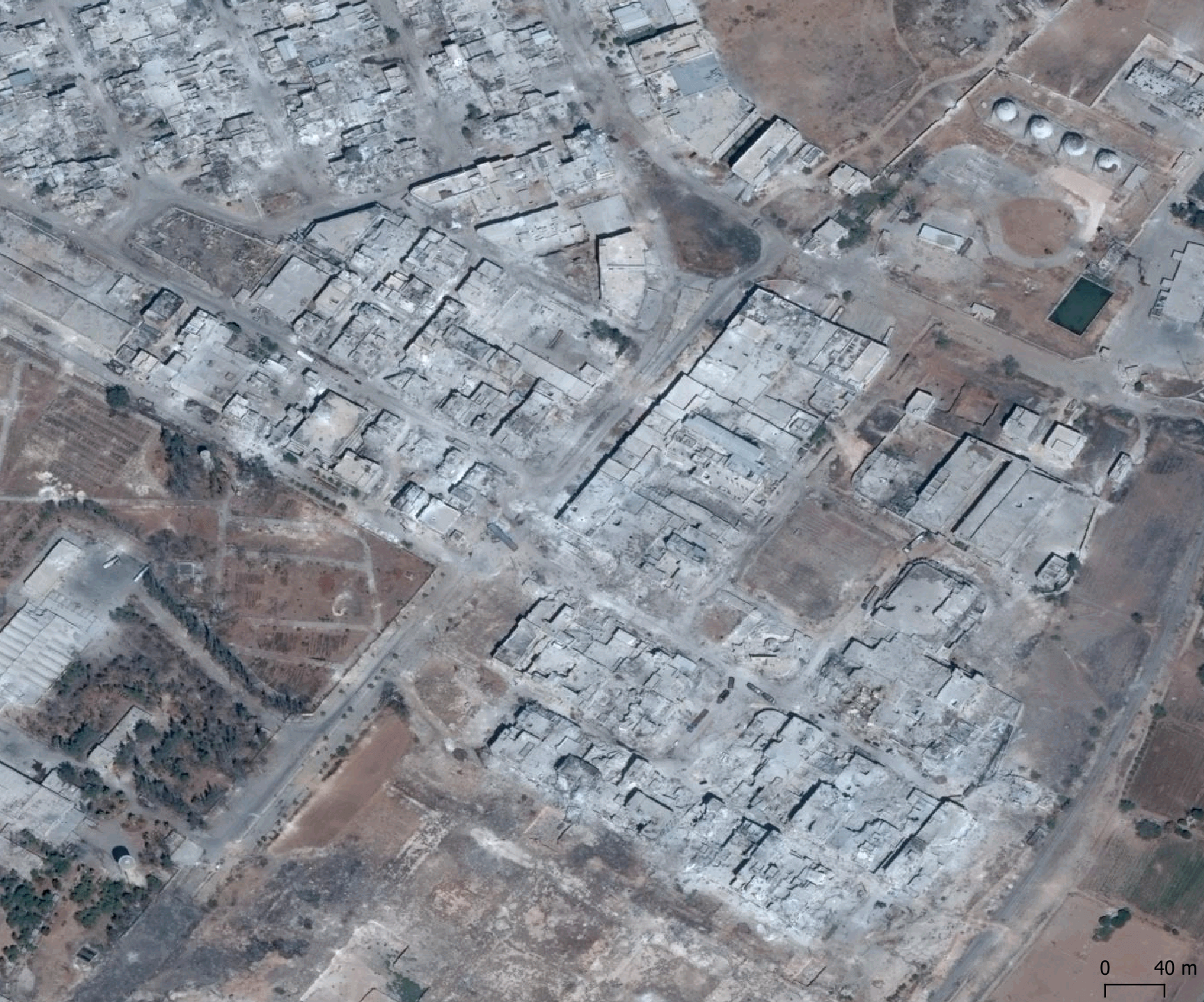}}
    }}
    \parbox{.33\figrasterwd}{%
      \subfloat[Continuous Prediction Scores 09/18/2016]{{
      \includegraphics[width=\hsize]{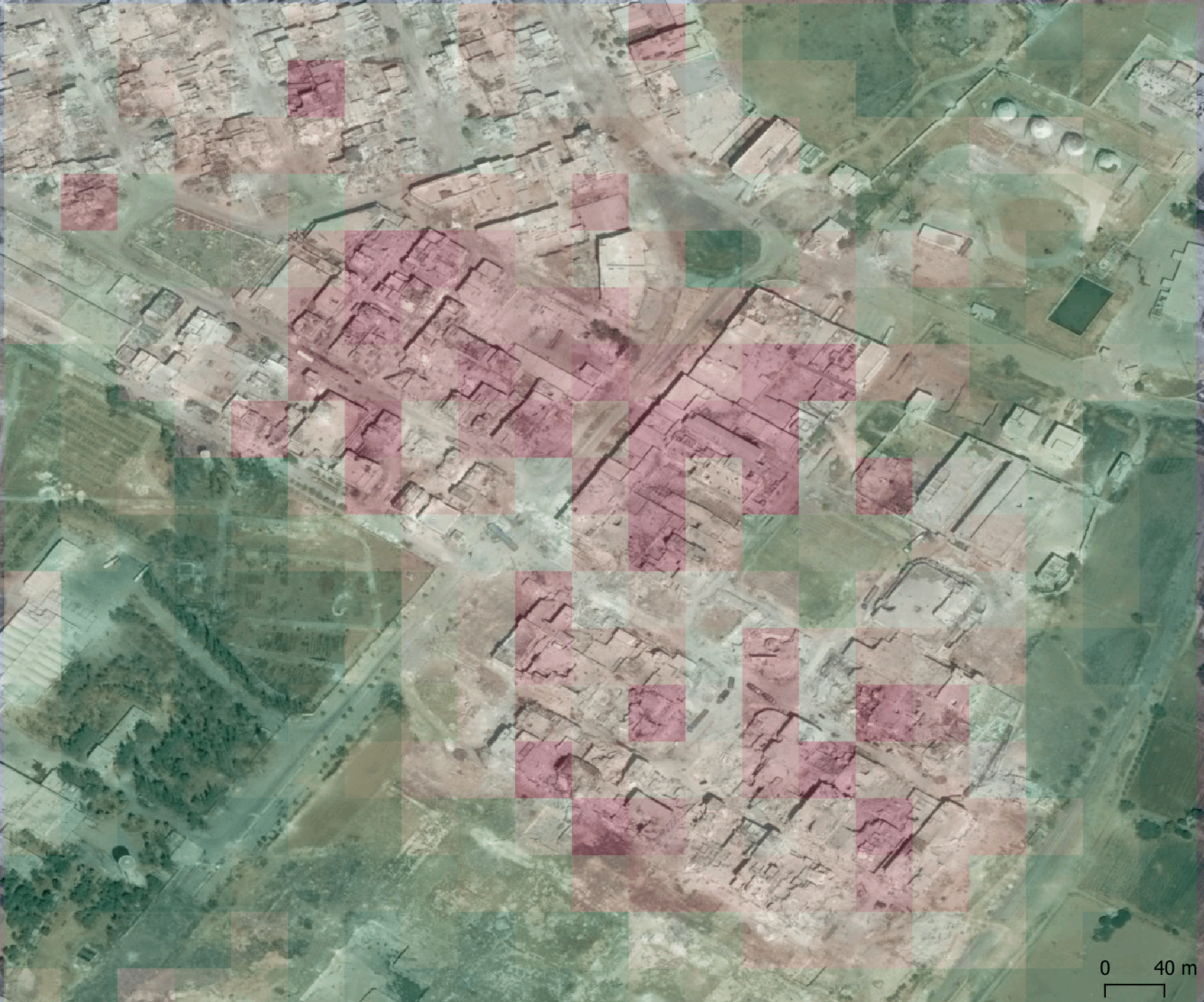}}
    }}
    \parbox{.33\figrasterwd}{%
      \subfloat[Binary Predictions 09/18/2016]{{
      \includegraphics[width=\hsize]{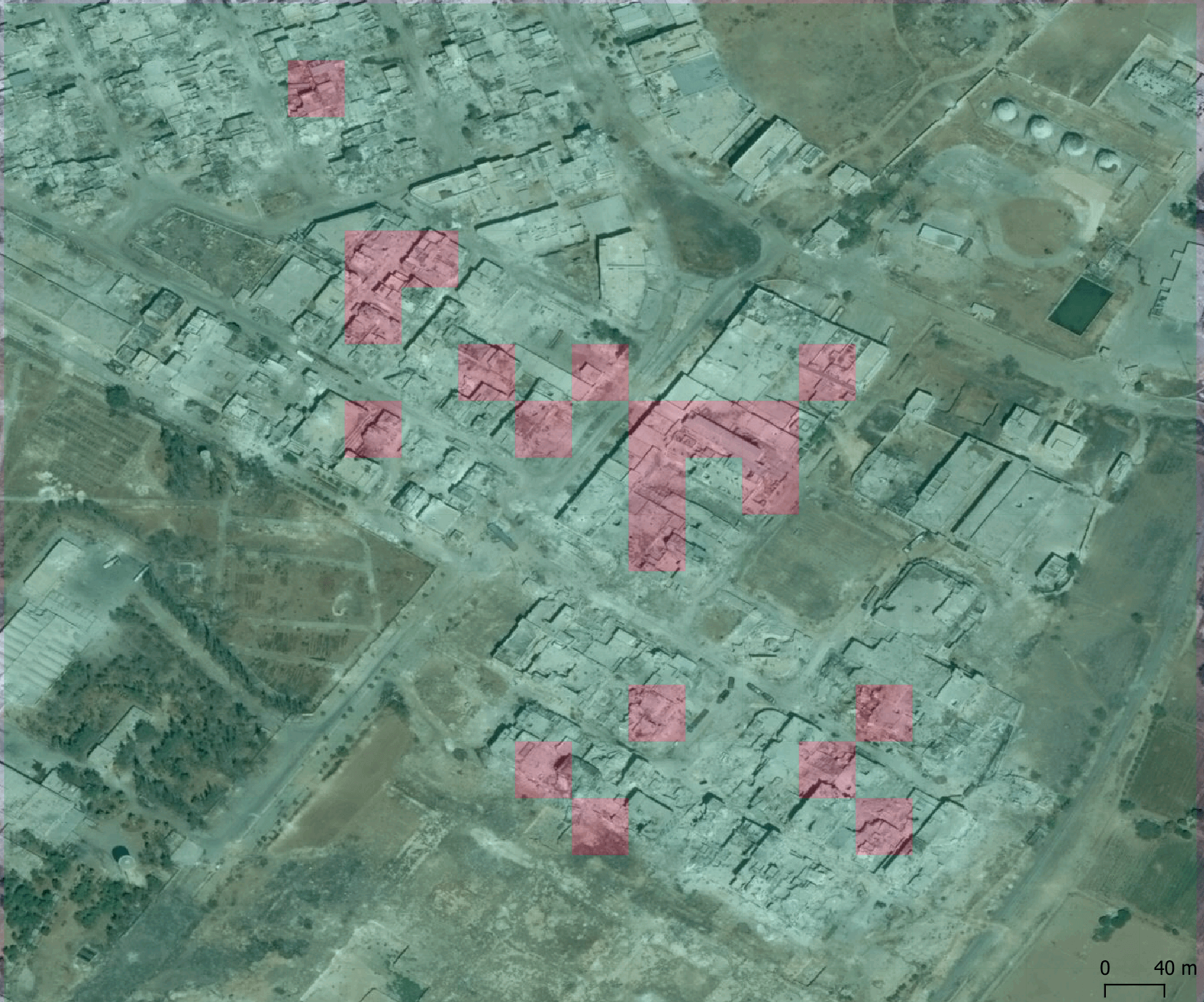}}
    }}
    }
\caption{Example of Raw Satellite Images (left panel) and Second-stage Patch-wise Continuous Predictions Scores (middle panel) and Binary Classification (right panel) for Ramouse Neighborhood of Aleppo, Syria, Before (first row) and After (second row) Heavy Weaponry Attacks. Red shade of patches indicates destruction. Satellite image recording dates: 06/12/2016 (before) and 09/18/2016 (after). Sources: Google Earth/Maxar Technologies and author calculations.}
\label{fig:exhibit_4}
\end{figure*}

Figure \ref{fig:exhibit_3} summarizes our main results for Aleppo. The top left panel (a) presents two precision/recall curves from the test sample which depict the out-of-sample performance of our classification approach. The first curve, in orange, plots the precision-recall trade-off in the balanced sample. The average precision here is 0.82. Precision in this sample declines slowly and smoothly for increasing levels of recall when moving to the right along the horizontal axis. Such performance suggests a very robust prediction with high performance. In contrast, the blue line depicts the performance of the same model when taking into account the unbalanced classes as in actual applications. Clearly precision is much lower with average precision being a mere 0.11. Precision is somewhat higher for very low levels of recall but never exceeds 0.4.

In the bottom left panel (b) we illustrate the improvement in precision that we achieve by applying the second stage of the method. The figure shows precision-recall curves from the dense prediction sample for the first stage (blue line) as in (a) and the improvement from the second stage (orange line), both evaluated in the unbalanced test sample. The second stage average precision improves significantly due to smoothing to 0.33 and, as depicted in panel (b), the slope of the precision-recall curve now represents a real trade-off when generating data. Low recall of just 0.3 would allow to reach precision levels close to 0.5. Alternatively, a cutoff which provides a recall of 0.5 leads to a precision level of just below 0.3 to produce a binary prediction from the second stage. The corresponding AUC is 0.90.

On the right panel (c) of Figure \ref{fig:exhibit_3} we show the continuous dense prediction scores coming out of the second stage for the 18\textsuperscript{th} of September 2016 across the entire city of Aleppo, including no-analysis zones. Red color indicates high predicted scores and green indicates low ones. Clearly, the red areas coincide with the labeled data in Figure \ref{fig:exhibit_1}. In addition, roads and parks are clearly visible as dark green (lowest probability) or yellow patches. This is not only evidence of the power of our approach in picking up housing destruction but it also shows how the CNN has learned that roads and parks are never destroyed buildings. This is a reassuring finding because it indicates that the approach is capable of identifying buildings without explicitly training this aspect.

\subsection{Validation Exercises}

We conduct several validation exercises to illustrate the merits of our approach. We first make use of the no-analysis areas in Aleppo (see Figure \ref{fig:exhibit_1}) that have been excluded from the training process. One of these zones corresponds to the Ramouse neighborhood in the southernmost tip of Aleppo -- an area which our classifier identified as heavily destroyed as depicted in panel (c) of Figure \ref{fig:exhibit_3}. Due to the classifier not having been trained on this areas, predicting destruction there is a good out-of-sample test.

Figure \ref{fig:exhibit_4} shows a close view into this area and also provides a timeline by showing both the raw images and predictions in June 2016 (top row) and in September 2016 (bottom row), respectively. In both rows we show the raw satellite picture (left), the continuous prediction scores coming out of the second stage (center) and a binary prediction generated to reach a recall of about 50 percent (right). The raw image in panel (a) shows that there was no visible destruction in this area in June, but that building destruction was dramatic by September, after a series of heavy weaponry attacks as reflected in panel (d). The center panels (b) and (e) show the same satellite images with an added layer depicting the continuous prediction scores, with green (red) indicating low (high) destruction probabilities. Comparing the evaluations before the attack (b) to the scores after the attack (e), shows a dramatic increase in predicted destruction by our approach which coincides clearly with the actual destruction of buildings in the area. Note that the model also classifies areas without buildings such as the industrial compound in the North, the fields and roads in the East, and the forest in the West of the image correctly as not destroyed.


\begin{figure}[h]
  \centering
  \includegraphics[width=10cm]{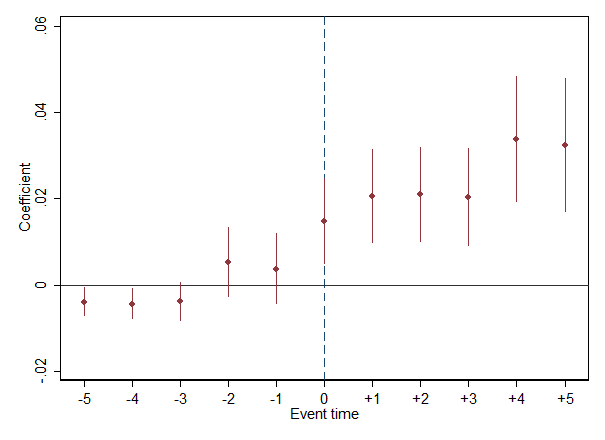}
\caption{Event Study Validation Exercise. External bomb event data from \emph{LiveUAmap} is positively and significantly correlated with satellite predicted war destruction at the patch level. The figure shows coefficients from a regression of 5 leads and lags of bombing events identified in the event data against our continuous destruction prediction score from the second stage. Dashed line indicates the occurrence of a bombing event in the event data and coefficients capture the response in predicted damage. The full regression specification and table can be found in Appendix Table S1. Sources: LiveUAmap event data and author calculations.}
\label{fig:exhibit_5}
\end{figure}


This event study demonstrates that the classifier is able to pick up destruction in parts of the city which were not part of the training sample. This is important as it shows that we are able to successfully solve the spatial and temporal domain bias problems within Aleppo and thus generate a consistent time series of destruction data this way. 

To verify this quantitatively for the entire sample of Aleppo we leverage a dataset of georeferenced violent events in Syria -- the Live Universal Awareness Map (\href{https://syria.liveuamap.com/}{LiveUAmap}) project. This database provides information on different types of conflict events during the Syrian civil war starting in 2015 -- including heavy weaponry events caused by bombs and artillery -- along with their geographical coordinates. We merge this data with our second stage results at the patch level to verify that our destruction predictions coincide with the location and timing of heavy weaponry attacks recorded in this event database. 

We conduct an event study where we rely on 528 distinct bombing events from the \textit{LiveUAmap} data for Aleppo and match them with the panel of destruction at the patch-level, which results in a sample of around 1.9 million observations. Note that this results in the ratio of events to images to be 528/1.9 million and receiving a signal from the timing of these events is therefore extremely ambitious. We regress the continuous destruction measure from the second stage on a dummy for the event data and a series of 5 leads and lags of all bombing events. In addition, we saturate the regression with fixed effects so that it filters out all variation at the cell level and all general changes across images. The coefficients in the event study represent the difference of the prediction score on patches that suffered a bombing event from those that did not. We present a coefficient summary plot in Figure \ref{fig:exhibit_5} which shows clearly that our remote sensing method is able to detect bombing events. The y-axis shows the coefficient size and the horizontal 0 line indicates the benchmark score during all periods leading up to the -5 period before the event. Scores increase significantly when an event takes place (at event time 0). Full regression output for this validation exercise is available in Table S1.


\subsection{Results Across Syrian Cities}
The method we propose here generalizes well across Syrian cities. To demonstrate this we collected images for five additional cities in Syria for which UNOSAT provides ground truth labels and sufficient Google Earth imagery is available (i.e. Daraa, Deir-Ez-Zor, Hama, Homs and Raqqa) and applied our method to the resulting sample of six cities. It is important to keep in mind that, while these cities are all in the same country, they are of different size, have different building types and are situated in different landscapes with a variety of vegetation and seasonal changes. If the neural network can adapt to these very different conditions it means we can be optimistic about applications elsewhere.

Table \ref{tab:table_1} summarizes the result. In the first column we report the total number of observations for each of the six cities, which is determined by city size and the number of dates for which images are available on the Google Earth archive throughout the study period (column 2). The vast majority of images in our sample comes from Aleppo due to its large size and high image availability - only around one third of all images come from other cities. We again exploit the time dimension of the labels to boost the training sample. Column three reports the resulting number of patches with attached destruction ground truth labels. The fourth column reports the share of destroyed patches and shows that destruction is present only in a small share of all samples. But there is also some interesting variation here with around 6 percent of the labeled patches from Homs being destroyed according to UNOSAT compared to around 1 percent for Daraa.


Columns five to seven of Table \ref{tab:table_1} report the performance in the test sample. The average AUC in the sample is 0.91 which is slightly higher than in the Aleppo-only sample. Performance in patches from Aleppo increases to 0.92 and the worst performance we have is from Deir-Ez-Zor with an AUC of 0.80. This is remarkable for two reasons. First, performance in the Aleppo sample increases slightly with adding samples from other cities to the training set. Second, despite the low share of training data from other cities we get good performance on the testing sample outside Aleppo.

Average precision in the balanced sample ranges between 0.78 in Daraa and 0.96 in Homs. Column seven reports precision in the actual, unbalanced sample (see Figure S5 for details). Here, the numbers vary much more, mostly as a function of the share of destroyed patches. Again this highlights the importance of the underlying balance in the sample, with our method reaching an average precision over 0.55 in Homs and Hama but only 0.12 in Daraa. This is a key take-away for applications: the generated data will not be precise in cities that face little or no destruction. In those cases it will be important to verify predictions with human input.

In the final two columns of Table \ref{tab:table_1} we contrast our binary destruction measure at the end of the respective sample in time (column eight) with the cumulative fatalities as identified by the Uppsala Conflict Data Program (UCDP) during the same period (column nine) \cite{Sundberg2013}. Obviously, both of these statistics are measured with some error. But to understand the potential of destruction data it is useful to see how much they differ across cities (See Figure S7). While in some cities destruction and fatalities occur proportionally (i.e. Aleppo, Daraa and Hama), in other cities they differ substantially with relatively low fatality figures despite high destruction. Most striking in this regard is Homs which, according to our estimates, suffered more destruction than Aleppo, but fewer fatalities.

These differences between our destruction data and the standard violence data suggest a research agenda on its own right. At what stage in a conflict is building destruction used? What can be done to reduce civilian fatalities during urban warfare? What are the effects of building destruction on displacement compared to other kinds of violence such as small fire arms? Can reporting-based violence data be used to reduce error in the remote sensing exercise? Can destruction data be used to reveal biases in reporting-based measures?


 \begin{table*}[t]
 \centering
 \small

 \caption{Performance Across Syrian Cities}\label{tab:table_1}
 \begin{tabular}{| c | c | c |c |c |c |c |c |c |c |} \hline \hline
 City & \thead{(1)\\\\Total\\samples}& \thead{(2)\\Number\\of dates/\\images} & \thead{(3)\\\\Labeled\\samples} & \thead{(4)\\\\Share\\destroyed} & \thead{(5)\\\\\\AUC} & \thead{(6)\\Average\\precision\\(1:1)} & \thead{(7)\\Average\\precision\\(unbal.)} & \thead{(8)\\\\Destruction\\binary} & \thead{(9)\\\\UCDP - GED\\fatalities} \\
 \midrule
 Aleppo &  \thead{2,106,412} &  \thead{22} & \thead{1,626,920} & \thead{0.02} & \thead{0.92} & \thead{0.90} & \thead{0.36} & \thead{12,585} & \thead{12,191} \\ \hline
 Daraa &  \thead{202,462} &  \thead{13} & \thead{125,231} & \thead{0.01} & \thead{0.89} & \thead{0.78} & \thead{0.12} & \thead{1,886} & \thead{2,270} \\ \hline
 Deir-Ez-Zor &  \thead{98,602} &  \thead{7} &  \thead{84,723} & \thead{0.03} & \thead{0.80} & \thead{0.83} & \thead{0.22} & \thead{6,553} & \thead{1,021} \\ \hline
 Hama &  \thead{285,057}&  \thead{9} &  \thead{224,365} & \thead{0.04} & \thead{0.91} & \thead{0.95} & \thead{0.68} & \thead{729}& \thead{318} \\ \hline
 Homs &  \thead{200,035}&  \thead{5} & \thead{83,941} & \thead{0.08} & \thead{0.86} & \thead{0.96} & \thead{0.56} & \thead{21,777}& \thead{1,414} \\ \hline
 Raqqa &  \thead{180,184}&  \thead{8} & \thead{112,481} & \thead{0.02} & \thead{0.88} & \thead{0.86} & \thead{0.32} & \thead{6,588}& \thead{1,502} \\ \hline \hline
 Total/average & \thead{3,072,752} & & \thead{2,257,661} & \thead{0.02} & \thead{0.91} & \thead{0.90} & \thead{0.38}&  \thead{}& \thead{} \\ \hline 
 \bottomrule
  \multicolumn{10}{l}{%
  \begin{minipage}{\textwidth}%
   \vspace{1.5mm}
   \textit{Sources}: Author calculations, UNITAR/UNOSAT damage annotations for Syria, and UCDP Georeferenced Event Dataset (GED) Global version 20.1. \textit{Note}: Performance metrics reflect post-second machine learning stage. Testing and training sets are defined at the patch/image location so that no location used for testing is also used in training. Training is balanced but not stratified by city so that Aleppo dominates the training sample. Column "destruction binary" is based on binary predictions from the second stage tuned to meet a 50 percent true positive rate at the patch level and refer to the last time period for each city. Column "UCDP-GED fatalities" are based on the best estimate of fatalities in each city during the last time period relying only on those observations with exact location and date information.
\end{minipage}%
 }\\

 \end{tabular}

 \end{table*}

\section{Discussion}
Building destruction due to heavy weapon attacks is a particularly salient form of war-related violence. Destruction is often used as a military strategy to displace population and is responsible for tremendous human suffering beyond the loss of life. Likewise organizations like the Red Cross warn that massive destruction of urban infrastructure (also called \textit{urbicide}) has dramatic knock-on effects on health as it implies the destruction of water and power supplies as well as hospitals. Therefore, reliable and updated data on destruction from war zones plays an important role for humanitarian relief efforts, but also for human rights monitoring, reconstruction initiatives, media reporting, as well as the study of violent conflict in academic research. Studying this form of violence quantitatively, beyond specific case studies, is currently impossible due to the absence of systematic data.

Our method of identifying building destruction combines the existing state of the art of machine vision methods with a novel postprocessing step, and exploits the time dimension of destruction data to expand the training data set. Thanks to these advances, we are able to achieve an AUC of 0.90 when applied to Aleppo, and an average precision of over 0.33 in the unbalanced sample. We show that our approach generalizes across different Syrian cities and is able to identify the timing and location of building destruction out-of-sample, i.e. in areas of Aleppo that have not been used for training the classifier. These results are extremely encouraging and allow applicability for automated destruction classification, and even close to real-time tracking for policy purposes.

However, our results also suggest limitations where average precision falls, e.g. if only a very low share, around one percent, of a city is destroyed. For applications requiring high precision in heavily imbalanced prediction problems such as the monitoring of several cities, we believe that the real use case for our approach will be in a decision support framework in which our results are combined with human verification to create much faster and accurate on-the-ground violence detection. Iterations between machine learning and human verification can also help improve the training process \cite{Colaresi17} and could be easily integrated in our methodology displayed in Figure \ref{fig:exhibit_1}. 

Our approach should be regarded as a first step in understanding the dynamic classification of building destruction over time. The classifier's ability to detect the location and timing of building damage also implies a potential for detecting the absence of building damage or reconstruction. An alternative application of our approach is, therefore, to augment the human-classification process of verifying violence -- so-called digital humanitarians \cite{meier2015digital} -- and to track the post-war recovery.

The performance of our approach could be further improved by increasing the size of the training dataset, which could also help adapting it to classify destruction in other war zones around the globe. Further performance improvement could be achieved through {\em fine tuning}, a common practice in deep learning in which the network is pre-trained with a large sample of building destruction from a variety of contexts in the first step, and then refined by training on heavy weaponry destruction. This could be implemented by using a recent public dataset of natural disaster destruction imagery that provides a sample of 98 thousand annotated buildings across 3 levels of damage \cite{Gupta_2019_CVPR_Workshops}. Moreover, domain adaptation techniques developed for deep learning could be used to try to further minimize the domain biases \cite{Csurka}. 

\vspace{8mm}

\textit{We would like to thank Joshua Blumenstock, Mathieu Couttenier, Joan Maria Esteban, Clément Gorin, Edward Miguel, and Sebastian Schütte for useful comments and discussions. We are grateful to Bruno Conte Leite, Jordi Llorens, Parsa Hassani, Dennis Hutschenreiter, Shima Nabiee, and Lavinia Piemontese for excellent research assistance. We are particularly grateful to Javier Mas for his research assistance which produced the coding backbone to this project. We thank seminar participants at University of California Berkeley, Empirical Studies of Conflict (ESOC) Project Annual Meeting, AI for Development conference by CEGA and World Bank Development Impact Evaluation Group (DIME), University of Bozen/Bolzano, International Institute of Social Studies (ISS) of Erasmus University, University of Economics Ho Chi Minh City, Lyon University, Trinity College Dublin, Barcelona GSE, IAE-CSIC, Universitat de Barcelona (UB), BeNA Winter Workshop/WZB Berlin, PREVIEW workshop at the German foreign office and ViEWS workshop in Uppsala. A.G. and H.M acknowledge financial support from the "la Caixa" Foundation project grant number CG-2017-04, title: "Analysing Conflict from Space", and from the Spanish Ministry of Science and Innovation, through the Severo Ochoa Programme for Centres of Excellence in R\&D (CEX2019-000915-S). H.M. acknowledges financial support from the Spanish Ministry of Science, Innovation and Universities through grant PGC-096133-B-100. A.G. also acknowledges financial support from the Spanish Ministry of Science, Innovation and Universities through grant PGC2018-094364-B-100. J.H. and A.M. acknowledge support from the Chapman University Faculty Opportunity Fund. Any remaining errors are our own.}


\bibliographystyle{unsrt}  

\bibliography{references.bib}

\end{document}